\documentclass[12pt]{article}

\usepackage[margin=1in]{geometry}
\usepackage{graphicx}
\usepackage{algorithm}
\usepackage[noend]{algpseudocode}
\usepackage{amsmath}
\usepackage{amssymb}
\usepackage{amsfonts}
\usepackage{latexsym}
\usepackage{pifont}
\usepackage{bbm}
\usepackage{hyperref}
\usepackage{mathrsfs}
\usepackage[nodisplayskipstretch]{setspace}
\usepackage{stackrel}
\usepackage{natbib}
\usepackage{xcolor}
\usepackage{appendix}
\usepackage{multirow}

\newcommand{\qed}{\nobreak \ifvmode \relax \else 
      \ifdim\lastskip<1.5em \hskip-\lastskip
      \hskip1.5em plus0em minus0.5em \fi \nobreak
      \vrule height0.75em width0.5em depth0.25em\fi}

\newcommand\blfootnote[1]{%
  \begingroup
  \renewcommand\thefootnote{}\footnote{#1}%
  \addtocounter{footnote}{-1}%   
  \endgroup
}

\begin{document}

\title{Review and Demonstration of a Mixture Representation for Simulation from Densities Involving Sums of Powers}
\author{Maryclare Griffin}
\date{\today}

\maketitle

\blfootnote{\raggedright \hspace{-2.15em}
$^1$Department of Mathematics and Statistics, University of Massachusetts Amherst, Amherst, MA 01003 (\href{mailto:maryclaregri@umass.edu}{\texttt{maryclaregri@umass.edu}}). \\
This research was supported by NSF grant DMS-2113079. Associated code is available at \href{https://github.com/maryclare/EPStan}{\texttt{https://github.com/maryclare/EPStan}}. The author gratefully acknowledges helpful comments from Hanyu Xiao.}

\begin{abstract} 
Penalized and robust regression, especially when approached from a Bayesian perspective, can involve the problem of simulating a random variable $\boldsymbol z$ from a posterior distribution that includes a term proportional to a sum of powers, $\|\boldsymbol z \|^q_q$, on the log scale. However, many popular gradient-based methods for Markov Chain Monte Carlo simulation from such posterior distributions use Hamiltonian Monte Carlo and accordingly require conditions on the differentiability of the unnormalized posterior distribution that do not hold when $q \leq 1$ \citep{Plummer2023}. This is limiting; the setting where $q \leq 1$ includes widely used sparsity inducing penalized regression models and heavy tailed robust regression models. In the special case where $q = 1$, a latent variable representation that facilitates simulation from such a posterior distribution is well known. However, the setting where $q < 1$ has not been treated as thoroughly. In this note, we review the availability of a latent variable representation described in \cite{Devroye2009}, show how it can be used to simulate from such posterior distributions when $0 < q < 2$, and demonstrate its utility in the context of estimating the parameters of a Bayesian penalized regression model.

\smallskip

\end{abstract}

\doublespacing
\newpage

% FYI - reference here for Laplace distribution in STAN: https://mc-stan.org/docs/functions-reference/unbounded_continuous_distributions.html#double-exponential-laplace-distribution

% To Do: 
% - Revise figure legends/text
% - Proofread multiple times

% Get to 10 pages

\section{Introduction}

Consider the problem of simulating a length-$n$ random variable $\boldsymbol z$ from a distribution with density $p(\boldsymbol z) \propto \text{exp}\left\{-f(\boldsymbol z) \right\}$.
Modern methods, specifically gradient-based methods such as Hamiltonian Monte Carlo (HMC), may either require or greatly benefit from the following: (i) differentiability of $f(\boldsymbol z)$ and (ii) availability of a closed form expression for $f(\boldsymbol z)$ \citep{Plummer2023, Strumbelj2024}.
Letting $g(\boldsymbol z)$ refer to the part of $f(\boldsymbol z)$ that is differentiable and partitioning $\boldsymbol z = (\boldsymbol z_1, \boldsymbol z_2)$ into two components $\boldsymbol z_1$ and $\boldsymbol z_2$ of length $n_1$ and $n_2$, a common form of $f(\boldsymbol z)$ that violates (i) is given by
\begin{align}\label{eq:naive}
f\left(\boldsymbol z \right) = g\left(\boldsymbol z\right) + \lambda \|\boldsymbol z_2\|^q_q
\end{align}
when $q \leq 1$. This arises in many settings including robust and penalized regression \citep{Poirier1986,Butler1990, Frank1993,Polson2014}. 

A well known way of addressing this problem in the case of $q = 1$ is to leverage the scale mixture representation of the density proportional to $\text{exp}\{-\lambda \|\boldsymbol z_2\|_1 \}$ \citep{West1987, Park2008, Hans2009, Ding2018}. Specifically, when $\boldsymbol z_2$ has a density proportional to $\text{exp}\{-\lambda \|\boldsymbol z_2\|_1 \}$, we can equivalently write that $z_{2i} |v_i \sim\text{normal}(0, v_i/\lambda)$ where $v_i \sim \text{exponential}(1/2)$. Accordingly, we can simulate $\boldsymbol z$ according to the density proportional to $\text{exp}\{-g\left(\boldsymbol z\right) - \lambda\|\boldsymbol z_2\|_1 \}$ by simulating $\boldsymbol z$ and $\boldsymbol v$ according to the density proportional to 
\begin{align*}
\text{exp}\left\{-g\left(\boldsymbol z\right) - \frac{1}{2} \left(\boldsymbol 1'\text{log}\left(\boldsymbol v\right)\right) - \frac{\lambda}{2}\left(\boldsymbol z_2'\text{diag}\left\{\boldsymbol v\right\}^{-1}\boldsymbol z_2\right) - \frac{1}{2}\left(\boldsymbol 1'\boldsymbol v\right) \right\}.
\end{align*}
This is helpful because $(\boldsymbol 1'\text{log}(\boldsymbol v))/2 + \lambda(\boldsymbol z_2'\text{diag}\{\boldsymbol v\}^{-1}\boldsymbol z_2)/2 + (\boldsymbol 1'\boldsymbol v)/2$ is a differentiable function of $\boldsymbol z_2$ and $\boldsymbol v$. This representation has been used extensively for Bayesian computation \citep{Park2008, Hans2009}.

When $q\neq 1$ and when $\boldsymbol z_2$ has a density proportional to $\text{exp}\{-\lambda \|\boldsymbol z_2\|_q^q \}$, the equivalent normal scale mixture representation is $z_{2i} |v_i \sim \text{normal}(0, v_i/\lambda^{2/q})$ where $v_i$ are independent and polynomially tilted positive $\alpha$-stable distribution with index of stability $\alpha = q/2$ \citep{West1987}. An analogous approach simulates $\boldsymbol z$ according to the density proportional to $\text{exp}\{-g(\boldsymbol z) - \lambda\|\boldsymbol z_2\|_q^q) \}$ by simulating $\boldsymbol z$ and $\boldsymbol v$ according to the density proportional to 
\begin{align*}
\text{exp}\left\{-g\left(\boldsymbol z\right) - \frac{1}{2} \left(\boldsymbol 1'\text{log}\left(\boldsymbol v\right)\right) - \frac{\lambda^{\frac{2}{q}}}{2}\left(\boldsymbol z_2'\text{diag}\left\{\boldsymbol v\right\}^{-1}\boldsymbol z_2\right) - \boldsymbol 1'h\left(\boldsymbol v; q\right)\right\},
\end{align*}
where $h(\boldsymbol v; q)$ is a function that does not have a closed form expression \citep{Polson2014}.

This can be resolved by further recognizing the distribution of the scales $v_i$ as a rate mixtures of generalized gamma random variables \citep{Devroye2009}. We can write polynomially titled positive $\alpha$-stable $v_i$ as equal in distribution to a transformation of gamma and Zolotarev distributed random variables $\xi_i > 0$ and $0 < \delta_i < \pi$,
\begin{align*}
&v_i\stackrel{d}{=} \frac{1}{2}\left(\xi_i^\frac{2-q}{q} \text{sin}\left(\left(\frac{q}{2}\right)\delta_i\right)^{-1} \text{sin}\left(\left(\frac{2 - q}{2}\right) \delta_i\right)^{\frac{q-2}{q}} \text{sin}\left(\delta_i\right)^{\frac{2}{q}} \right)\text{, } \\
&\xi_i \stackrel{i.i.d.}{\sim} \text{gamma}\left(\text{shape}=\frac{2 + q}{2q}, \text{rate}=1\right)\text{ and } \\
&p\left(\delta_i | q\right)= \left(\frac{\Gamma\left(1 + \frac{1}{2}\right)\Gamma\left(\frac{1}{2} + \frac{1}{q}\right)}{\pi \Gamma\left(1 + \frac{1}{q}\right)}\right) \text{sin}\left(\left(\frac{q}{2}\right) \delta_i\right)^{-\frac{1}{2}}\text{sin}\left(\left(\frac{2 - q}{2}\right)\delta_i\right)^{\frac{q-2}{2q}}\text{sin}\left(\delta_i\right)^{\frac{1}{q}}.
\end{align*}
This implies that we can simulate $\boldsymbol z$ according to the density proportional to $\text{exp}\{-g(\boldsymbol z) - \lambda\|\boldsymbol z_2\|^q_q\}$ by simulating $\boldsymbol z$, $\boldsymbol \xi$, and $\boldsymbol \delta$ according to the density proportional to  
\footnotesize
\begin{align}\label{eq:nsm}
&\text{exp}\left\{-g\left(\boldsymbol z\right) - \lambda^{\frac{2}{q}}\left(\boldsymbol 1'\left(\boldsymbol \xi^{\frac{q - 2}{q}} \circ \text{sin}\left(\left(\frac{q}{2}\right)\boldsymbol \delta\right)\circ \text{sin}\left(\left(\frac{2 - q}{2}\right)\boldsymbol \delta\right)^{\frac{2-q}{q}} \circ \text{sin}\left(\boldsymbol \delta\right)^{-\frac{2}{q}} \circ \boldsymbol z_2^2\right)\right) - \boldsymbol 1'\boldsymbol \xi\right\},
\end{align}
\normalsize
where `$\circ$' refers to the elementwise Hadamard product and $\text{sin}\left(\cdot\right)$ and $\left(\cdot\right)^a$ are applied elementwise. 
This is differentiable with respect to $\boldsymbol z$, $\boldsymbol \xi$, and $\boldsymbol \delta$. A derivation which includes the normalizing constant $C(q, \lambda) = \int \text{exp}\{-\lambda \|\boldsymbol z_q\|^q_q\} d \boldsymbol z_2$ is provided in an Appendix. 

In some situations, alternative ``non-centered'' parametrization may be preferable \citep{Betancourt2015}. We can simulate $\boldsymbol z$ according to the density proportional to $\text{exp}\{-g(\boldsymbol z) - \lambda\|\boldsymbol z_2\|^q_q\}$ by simulating $\boldsymbol z_1$, $\boldsymbol w$, $\boldsymbol \xi$, and $\boldsymbol \delta$ according to the density proportional to  
\footnotesize
\begin{align}\label{eq:prod}
\text{exp}\left\{\right.&-g\left(\left(\boldsymbol z_1, 2^{-\frac{1}{2}} \lambda^{-\frac{1}{q}}\left(\boldsymbol \xi^{\frac{2-q}{2q}} \circ\text{sin}\left(\left(\frac{q}{2}\right)\boldsymbol \delta\right)^{-\frac{1}{2}}\circ\text{sin}\left(\left(\frac{2 - q}{2}\right)\boldsymbol \delta\right)^{\frac{q-2}{2q}}\circ\text{sin}\left(\boldsymbol \delta\right)^{\frac{1}{q}}  \circ \boldsymbol w\right)\right)\right) +\\ \nonumber
&- \left(\frac{1}{2}\right)\boldsymbol w'\boldsymbol w + \left(\frac{2 - q}{2q}\right)\boldsymbol 1'\text{log}\left(\boldsymbol \xi\right) - \boldsymbol 1'\boldsymbol \xi  + \left(\frac{q-2}{2q}\right)\boldsymbol 1'\text{log}\left(\text{sin}\left(\left(\frac{2 - q}{2}\right)\boldsymbol \delta\right)\right) + \\ \nonumber
&\left.   - \left(\frac{1}{2}\right)\boldsymbol 1'\text{log}\left( \text{sin}\left(\left(\frac{q}{2}\right)\boldsymbol \delta\right)\right) + \left(\frac{1}{q}\right)\boldsymbol 1'\text{log}\left(\text{sin}\left(\boldsymbol \delta\right)\right) \right\},
\end{align}
\normalsize
and setting $\boldsymbol z_2 = 2^{-1/2} \lambda^{-1/q} (\boldsymbol \xi^{\frac{2-q}{2q}} \circ \text{sin}(q\boldsymbol \delta/2)^{-1/2}\circ \text{sin}((2 - q)\boldsymbol \delta/2)^{(q-2)/(2q)}\circ \text{sin}(\boldsymbol \delta)^{1/q} \circ \boldsymbol w)$.
Analogously, this is differentiable with respect to $\boldsymbol z_1$, $\boldsymbol w$, $\boldsymbol \xi$, and $\boldsymbol \delta$. As before,  derivation which includes the normalizing constant is provided in an Appendix.

\section{Demonstration}

We demonstrate the use of these representations for simulation of regression coefficients $\boldsymbol z_2$ under a penalized regression model  relating an $m\times 1$ response $\boldsymbol y$ to an $m\times n_2$ matrix of covariates $\boldsymbol X$ via unknown parameters  $\boldsymbol \theta  = (\sigma^2, \lambda, q)$,
\begin{align}
    \boldsymbol y &= \boldsymbol X \boldsymbol z_2 + \boldsymbol e\text{, \quad} p\left(\boldsymbol z_2\right) \propto \text{exp}\left\{-\lambda \|\boldsymbol z_2\|^q_q \right\}\text{, \quad} \boldsymbol e \sim \text{normal}\left(\boldsymbol 0, \sigma^2 \boldsymbol I_m\right), \label{eq:penalized}
\end{align}
where $\boldsymbol I_m$ refers to an $m\times m$ identity matrix. When $\boldsymbol \theta$ is fixed, the penalized regression model \eqref{eq:penalized} has $g\left(\boldsymbol z\right) = -\| \boldsymbol y - \boldsymbol X \boldsymbol z_2\|^2_2/(2\sigma^2)$ and $\boldsymbol z = \boldsymbol z_2$.  

We consider two datasets that are frequently used in the relevant literature, which we refer to as the prostate and glucose data, respectively. Because the performance of algorithms for simulating from simulating a random variable is known to depend on the dimension of the random variable, the two datasets are chosen to exemplify simulation of a relatively low dimensional random variable, with dimension $n_2 = 8$, and a relatively high dimensional random variable, with dimension $n_2 = 72$. The prostate data has appeared in  \cite{Tibshirani1996} and contains measurements of log prostate specific antigen and $n_2 = 8$ clinical measures associated with prostate cancer progression for $m = 97$ subjects. The glucose data has appeared in \cite{Priami2015} and contains measurements of blood glucose concentration and $n_2 = 72$ metabolite measurements and health indicators for $m = 68$ subjects. 

We consider 9 values of $\boldsymbol \theta = (\sigma^2, \lambda, q)$ to compare the performance of methods for simulating from the posterior distribution of $\boldsymbol z_2$, denoted by $\boldsymbol \theta^{(1)}, \dots, \boldsymbol \theta^{(9)}$. We estimate some components of each $\boldsymbol \theta^{(k)}$ from the data and systematically vary others. We obtain estimates $\hat{\sigma}^2$ and $\hat{\tau}^2$ of the noise variance and the variance of the regression coefficients $\boldsymbol z_2$ by minimizing
\begin{align*}
\text{log}\left(\left|\boldsymbol X \boldsymbol X'\tau^2 + \boldsymbol I\sigma^2 \right|\right)/2 + \boldsymbol y'\left(\boldsymbol X \boldsymbol X'\tau^2 + \boldsymbol I\sigma^2\right)^{-1}\boldsymbol y
\end{align*}
with respect to $\sigma^2$ and $\tau^2$. We set $\theta^{(k)}_1 = \hat{\sigma}^2$, $\theta^{(k)}_2 = (\Gamma(3/\theta^{(k)}_3)/(\hat{\tau}^2\Gamma(1/\theta^{(k)}_3)))^{\theta^{\left(k\right)}/2}$, and $\theta^{(k)}_3 = 2k/10$, which fixes the prior variance of $\boldsymbol z_2$ to $\hat{\tau}^2$ as in \citep{Griffin2017c}.

For each $\boldsymbol \theta^{(k)}$, we simulate from the posterior distribution of $\boldsymbol z_2$ under the model described by Equation~\eqref{eq:penalized} using STAN without changing the default settings \citep{Carpenter2017}. We compare posterior simulation using Representation~\eqref{eq:nsm} which we refer to as the \emph{centered} parametrization, posterior simulation using Representation~\ref{eq:prod} which we refer to as the \emph{non-centered} parametrization, and posterior simulation directly from \eqref{eq:naive} which we refer to as the \emph{naive} parametrization. For each, we obtain 10 chains. Each chain simulates  1,000 burn-in (warmup) iterations followed by 1,000 draws which are retained. Starting values for $\boldsymbol z_2$ are shared across methods, i.e. the first chain for all three posterior simulation methods for the same data and parameters $\boldsymbol \theta^{(k)}$ shares the same starting value for $\boldsymbol z_2$.

\begin{figure}[h!]
    \centering
    \includegraphics{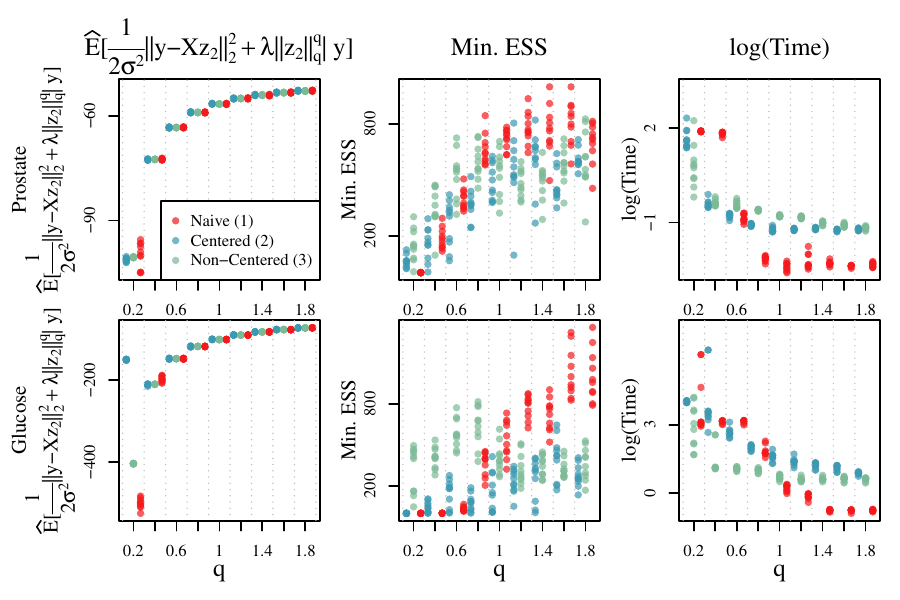}
    \caption{Estimated posterior mean log unnormalized posterior $\hat{E}[\|\boldsymbol y - \boldsymbol X \boldsymbol z_2 \|^2_2/(2\sigma^2) + \lambda \|\boldsymbol z_2 \|^q_q | \boldsymbol y]$, minimum effective sample size over all simulated parameters across 1,000 draws per chain after burn-in, and total time elapsed per chain for the centered parametrization~\eqref{eq:nsm}, the non-centered parametrization~\ref{eq:prod}, and the naive parametrization~\eqref{eq:naive} computed from the prostate and glucose datasets.}
    \label{fig:summary}
\end{figure}

First, we compare estimates of the posterior mean of the log unnormalized posterior $\hat{E}[\|\boldsymbol y - \boldsymbol X \boldsymbol z_2 \|^2_2/(2\sigma^2) + \lambda \|\boldsymbol z_2 \|^q_q | \boldsymbol y]$ based on the 1,000 simulated values retained after burn-in for each chain. These are shown in the first column of Figure~\ref{fig:summary}. Parametrizations that correspond to better simulation from the posterior are expected to produce estimates that are less variable across chains. All three parametrizations produce estimates that are very consistent across chains when $q \geq 0.8$. When $q < 0.8$ and $n_2$ is relatively small, the naive and centered parametrizations \eqref{eq:naive} and \eqref{eq:nsm} produce more variable estimates of the log unnormalized posterior across chains, moreso for smaller values of $q$. When $q < 0.8$ and $n_2$ is relatively large, the naive and centered parametrizations \eqref{eq:naive} and \eqref{eq:nsm} not only produce more variable estimates of the log unnormalized posterior across chains for some values of $q$ but also produce very different estimates of the log unnormalized posterior for others when different parametrizations are used.

The second column shows minimum effective sample sizes based on the 1,000 simulated values retained after burn-in for each chain. It suggests that the naive and centered parametrizations \eqref{eq:naive} and \eqref{eq:nsm} tend to provide smaller effective sample sizes and poorer simulation from the posterior when $q \leq 0.8$. The performance of the naive and centered parametrizations \eqref{eq:naive} and \eqref{eq:nsm} deteriorates as $q$ decreases, regardless of the dimension. In particular, the minimum effective sample sizes are nearly zero when the naive or centered parametrization \eqref{eq:naive} or \eqref{eq:nsm} are used. This indicates that the discordant estimates of the log  unnormalized  posterior across parametrizations observed for $q = 0.2$ and $n_2 = 64$ reflect that the estimates produced by the naive and centered parametrizations \eqref{eq:naive} and \eqref{eq:nsm} are incorrect, despite being precise. Unsurprisingly, we observe that the naive parametrization \eqref{eq:naive} tends to outperform the others when $q > 1$, moreso as $q$ increases and moreso when the dimension $n_2$ is greater. Interestingly, when $n_2 = 8$, the naive parametrization \eqref{eq:naive} performs competitively for $q = 0.8$ and tends to perform best when $q = 1$. This is not the case when the dimension $n_2$ is larger, and suggesting that simulating from a log posterior that is not differentiable using the naive parametrization \eqref{eq:naive} may be more feasible when the dimension is smaller and/or when the differentiable part $g\left(\boldsymbol z\right)$ of the unnormalized log posterior $f\left(\boldsymbol z\right)$ dominates more.

Last, the third column of Figure~\ref{fig:summary} shows that larger effective sample sizes are not costly to obtain in terms of computation time; the parametrizations that produce the largest minimum effective sample sizes tend to have the fastest run times.
The results shown in Figure~\ref{fig:summary} for $q > 1$ also highlight the cost of introducing $2\times n_2$ additional auxiliary random variables. This adds both time and reduces efficiency, as posterior simulation with auxiliary random variables tends to take longer and produce lower effective sample sizes given the same number of simulated values.

\begin{figure}[h!]
    \centering
    \includegraphics{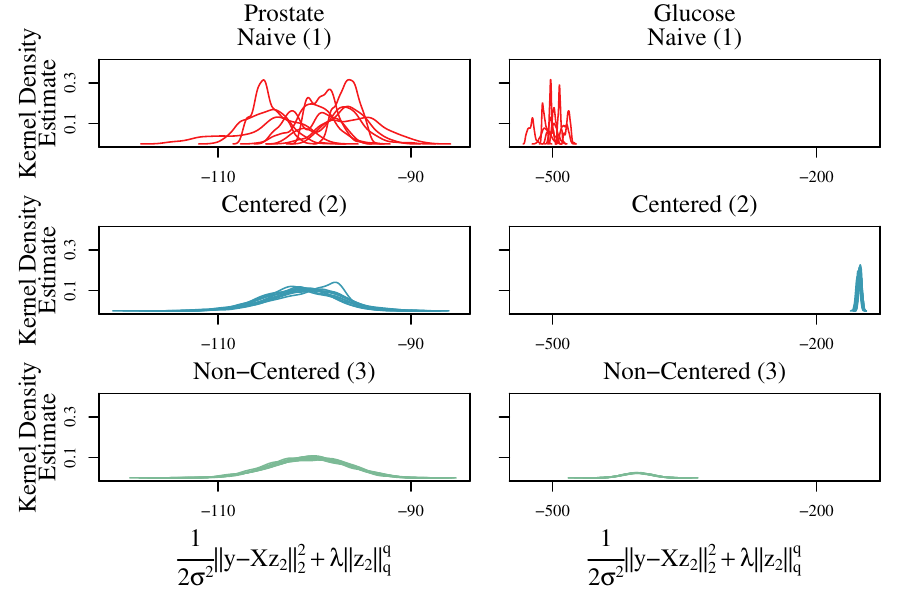}
    \caption{Kernel density estimates of the posterior density of the log unnormalized posterior $\boldsymbol y - \boldsymbol X \boldsymbol z_2 \|^2_2/(2\sigma^2) + \lambda \|\boldsymbol z_2 \|^q_q$ across chains for $q = 0.2$ based on 1,000 draws per chain after burn-in for the centered parametrization~\eqref{eq:nsm}, the non-centered parametrization~\ref{eq:prod}, and the naive parametrization~\eqref{eq:naive}  computed from the prostate and glucose datasets.}
    \label{fig:kdens}
\end{figure}

Figure~\ref{fig:kdens} helps us better understand what is happening when $q = 0.2$ for both datasets by showing kernel density estimates of the unnormalized log posterior $\|\boldsymbol y - \boldsymbol X \boldsymbol z_2 \|^2_2/(2\sigma^2) + \lambda \|\boldsymbol z_2 \|^q_q$ computed using the 1,000 simulated values of $\boldsymbol z_2$ retained after burn-in for each chain. Based on the high effective sample sizes and consistency across chains of the non-centered parametrization \eqref{eq:prod} observed in Figure~\ref{fig:summary}, we treat the kernel density estimates obtained by using the non-centered parametrization \eqref{eq:prod} as a benchmark or gold standard for both datasets. This is further supported by the similarity of kernel density estimates obtained from using the non-centered parametrization \eqref{eq:prod} across chains. When the dimension $n_2$ is relatively small, we see that all kernel density estimates share similar supports. However, the chains obtained using the naive parametrization \eqref{eq:naive} are very heterogenous, with distinct modes and shapes which are not consistent with the kernel density estimates obtained from the non-centered parametrization \eqref{eq:prod}. In contrast, the majority of chains obtained using centered parametrization \eqref{eq:nsm} are consistent with the results obtained by using the non-centered parametrization \eqref{eq:nsm}, but one chain appears to get stuck at another incorrect mode. 
When the dimension of $\boldsymbol z_2$ is relatively large, the naive and centered parametrizations \eqref{eq:naive} and \eqref{eq:nsm} do not share the same support as the non-centered parametrization \eqref{eq:prod}. Both appear to get stuck far from the mode of the posterior distribution of $\|\boldsymbol y - \boldsymbol X \boldsymbol z_2 \|^2_2/(2\sigma^2) + \lambda \|\boldsymbol z_2 \|^q_q$.

\begin{figure}[h!]
    \centering
    \includegraphics{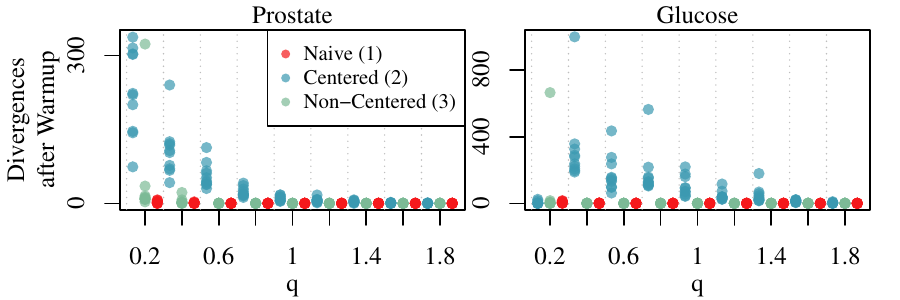}
    \caption{Divergent transitions after burn-in per chain for the centered parametrization~\eqref{eq:nsm}, the non-centered parametrization~\ref{eq:prod}, and the naive parametrization~\eqref{eq:naive}  computed from the prostate and glucose datasets.} 
    \label{fig:div}
\end{figure}

Last, we examine divergent transitions after burn-in, which indicate numerical instability associated with draws from the posterior that can be interpreted as evidence of poor simulation performance \citep{Carpenter2017}.  Divergences are prevalent for the centered parametrization \eqref{eq:nsm}, especially when $q$ is small or the dimension of $\boldsymbol z_2$ is large. This is consistent with the performance of centered parametrizations in other settings \citep{Betancourt2015}. Interestingly, the naive parametrization \eqref{eq:naive} does not yield divergent transitions after warmup, even when it produces low effective sample sizes and poor estimates. This highlights the need to exercise care when using gradient-based methods to simulate from posterior distributions that are not differentiable.

\section{Conclusion}

In this note, we demonstrate the use of a latent variable representations of posterior distributions for random $\boldsymbol z$ with densities proportional to $\text{exp}\left\{f\left(\boldsymbol z \right) = g\left(\boldsymbol z\right) + \lambda \|\boldsymbol z_2\|^q_q\right\}$. This is of great value as it expands access to previously difficult to use models for users of STAN \citep{Carpenter2017}, PyMC \citep{Salvatier2016}, and other software for Hamiltonian Monte-Carlo based posterior simulation as described in \cite{Strumbelj2024}. This includes bridge penalized linear and generalized linear regression models, robust linear models of the form
\begin{align*}
    \boldsymbol y &= \boldsymbol X \beta + \boldsymbol z_2\text{, \quad} p\left(\boldsymbol z_2\right) \propto \text{exp}\left\{-\lambda \|\boldsymbol z_2\|^q_q \right\},
\end{align*}
and models that use the structured shrinkage priors introduced \cite{Griffin2018}.

\bibliography{library.bib}
\bibliographystyle{chicago}

\begin{appendix}

\section*{Appendix}

Let
\begin{align*}
k\left(\delta_i | q\right)^{\frac{q-2}{2q}} &= 
\left(\text{sin}\left(\frac{q \delta_i}{2}\right)^{-\frac{q}{q - 2}} \text{sin}\left(\frac{(2 - q)\delta_i}{2}\right) \text{sin}\left(\delta_i\right)^{\frac{2}{q - 2}}\right)^{\frac{q-2}{2q}} \\
&= 
\text{sin}\left(\left(\frac{q}{2}\right)\delta_i\right)^{-\frac{1}{2}}\text{sin}\left(\left(\frac{2 - q}{2}\right)\delta_i\right)^{\frac{q-2}{2q}} \text{sin}\left(\delta_i\right)^{\frac{1}{q}}. 
\end{align*}

If $p\left(\boldsymbol z_2\right) = \frac{q}{2\Gamma\left(1/q\right)\lambda^{-1/q}}\text{exp}\left\{-\lambda \left|\left|\boldsymbol z_2\right|\right|^q_q \right\}$, then

\begin{align*}
    p\left(\boldsymbol z_2 | \boldsymbol \xi\right)p\left(\boldsymbol \xi | \boldsymbol \delta \right) p\left(\boldsymbol \delta\right) = \prod_{j = 1}^{n_2}& \left(\frac{\pi \left(\frac{\xi_i}{k\left(\delta_i|q\right)}\right)^\frac{2-q}{q}}{\lambda^\frac{2}{q}}\right)^{-\frac{1}{2}} \text{exp}\left\{-\frac{z_{2j}^2\lambda^\frac{2}{q}}{\left(\frac{\xi_i}{k\left(\delta_i|q\right)}\right)^\frac{2-q}{q}}\right\} \left(\frac{1}{\Gamma\left(\frac{2 + q}{2q}\right)} \right)\\
    &\left(\xi_i \right)^{\frac{2 + q}{2q} - 1}\text{exp}\left\{ -\xi_i \right\} \\
    &\left(\frac{\Gamma\left(1 + \frac{1}{2}\right)\Gamma\left(\frac{1}{2} + \frac{1}{q}\right)}{\pi \Gamma\left(1 + \frac{1}{q}\right)}\right) k\left(\delta_i | q\right)^{\frac{q - 2}{2q}} \\
    =\prod_{j = 1}^{n_2}& \left(\frac{1}{2\pi}\right)\left(\frac{q\lambda^{\frac{1}{q}}}{\Gamma\left(\frac{1}{q}\right)}\right) \times \\
    &\text{exp}\left\{ -\xi_i - \lambda^\frac{2}{q}z_{2j}^2 \xi_i^\frac{q - 2}{q}\text{sin}\left(\left(\frac{q}{2}\right)\delta_i\right)^{-1} \text{sin}\left(\left(\frac{2 - q}{2}\right)\delta_i\right)^{\frac{q-2}{q}}\text{sin}\left(\delta_i\right)^{\frac{2}{q}}\right\},
\end{align*}
where   $\Gamma(1 + 1/q) = \Gamma(1/q)/q$ and $\Gamma(3/2) = \pi^{1/2}/2$ follow from properties of the gamma function as described in Abramowitz and Stegun.

An alternative, set $\boldsymbol z_2 = \left(\frac{\boldsymbol \xi^{\frac{2-q}{2q}}}{\sqrt{2}\lambda^{\frac{1}{q}}k\left(\boldsymbol \delta|q\right)^{\frac{2-q}{2q}}}\right)\boldsymbol w$ and simulate from:
\begin{align*}
    p\left(\boldsymbol w_2 | \boldsymbol \xi\right)p\left(\boldsymbol \xi | \boldsymbol \delta \right) p\left(\boldsymbol \delta\right) = \prod_{j = 1}^{n_2}& \left(2\pi \right)^{-\frac{1}{2}} \text{exp}\left\{-\frac{w_{j}^2}{2}\right\} \left(\frac{1}{\Gamma\left(\frac{2 + q}{2q}\right)} \right)\\
    &\xi_i^{\frac{2 + q}{2q} - 1}\text{exp}\left\{ -\xi_i \right\} \\
    &\left(\frac{\Gamma\left(1 + \frac{1}{2}\right)\Gamma\left(\frac{1}{2} + \frac{1}{q}\right)}{\pi \Gamma\left(1 + \frac{1}{q}\right)}\right) k\left(\delta_i | q\right)^{\frac{q - 2}{2q}} \\
    = \prod_{j = 1}^{n_2}& \left(\frac{1}{2^{\frac{3}{2}}\pi}\right) \left(\frac{q}{\Gamma\left(\frac{1}{q}\right)}\right) \text{exp}\left\{-\frac{w_{j}^2}{2}\right\} \\
    &\xi_i^{\frac{2 - q}{2q}}\text{exp}\left\{ -\xi_i \right\} \text{sin}\left(\left(\frac{q}{2}\right)\delta_i\right)^{-\frac{1}{2}} \text{sin}\left(\left(\frac{2 - q}{2}\right)\delta_i\right)^{\frac{q-2}{2q}}\text{sin}\left(\delta_i\right)^{\frac{1}{q}} 
\end{align*}
where $\Gamma(1 + 1/q) = \Gamma(1/q)/q$ and $\Gamma(3/2) = \pi^{1/2}/2$ follow from properties of the gamma function as described in Abramowitz and Stegun.

\end{appendix}

\end{document}